\PassOptionsToPackage{table,xcdraw}{xcolor}
\documentclass[acmlarge]{acmart}
\usepackage{booktabs}
\usepackage{multirow}  
\usepackage{soul}
\usepackage{subfigure}
\usepackage{graphicx}
\usepackage{lipsum}
\usepackage{textcomp}
\usepackage{xcolor}
\definecolor{blue}{rgb}{0,0,0}
\AtBeginDocument{%
  \providecommand\BibTeX{{%
    \normalfont B\kern-0.5em{\scshape i\kern-0.25em b}\kern-0.8em\TeX}}}

\setcopyright{acmcopyright}
\acmJournal{IMWUT}
\acmYear{2022} \acmVolume{6} \acmNumber{3} \acmArticle{115}
\acmMonth{9} \acmPrice{15.00}\acmDOI{10.1145/3550335}

\begin{document}


\title{Individual and Group-wise Classroom Seating Experience: Effects on Student Engagement in Different Courses}


\author{Nan Gao}
\email{nannn.gao@gmail.com}
\affiliation{%
  \institution{RMIT University}
  \city{Melbourne}
  \country{Australia}
  \postcode{3000}
}
\author{Mohammad Saiedur Rahaman}
\email{saiedur.rahaman@rmit.edu.au}
\affiliation{%
  \institution{RMIT University}
  \city{Melbourne}
  \country{Australia}
  \postcode{3000}
}
\author{Wei Shao}
\email{weishao@ucdavis.edu}
\affiliation{%
  \institution{UC Davis}
  \city{Davis}
  \country{United States}
  \postcode{95616}
}
\author{Kaixin Ji}
\email{kaixin.ji@student.rmit.edu.au}
\affiliation{%
  \institution{RMIT University}
  \city{Melbourne}
  \country{Australia}
  \postcode{3000}
}
\author{Flora D. Salim}
\email{flora.salim@unsw.edu.au}
\affiliation{%
  \institution{University of New South Wales (UNSW)}
  \city{Sydney}
  \country{Australia}
  \postcode{1466}
}
\renewcommand{\shortauthors}{Gao et al.}

\begin{abstract}
Seating location in the classroom can affect student engagement, attention and academic performance by providing better visibility, improved movement, and participation in discussions. Existing studies typically explore how traditional seating arrangements (e.g. grouped tables or traditional rows) influence students' perceived engagement, without considering group seating behaviours under more flexible seating arrangements. Furthermore, survey-based measures of student engagement are prone to subjectivity and various response bias. Therefore, in this research, we investigate how individual and group-wise classroom seating experiences affect student engagement using wearable physiological sensors. We conducted a field study at a high school and collected survey and wearable data from 23 students in 10 courses over four weeks. We aim to answer the following research questions: \textit{1. How does the seating proximity between students relate to their perceived learning engagement? 2. How do students' group seating behaviours relate to their physiologically-based measures of engagement (i.e. physiological arousal and physiological synchrony)?} 
Experiment results indicate that the individual and group-wise classroom seating experience is associated with perceived student engagement and physiologically-based engagement measured from electrodermal activity. We also find that students who sit close together are more likely to have similar learning engagement and tend to have high physiological synchrony. This research opens up opportunities to explore the implications of flexible seating arrangements and has great potential to maximize student engagement by suggesting intelligent seating choices in the future.

\end{abstract}



\maketitle

\section{Introduction}

Anecdotal evidence has suggested that student seating experience has played an important role in affecting student engagement, participation, attention and academic performance in classrooms \cite{fernandes2011does,burda1996college}. 
Different seating locations provide different degrees of access to learning conditions and resources (e.g. the ability to see and hear the teacher \cite{benedict2004seating}), which affects students' physical and mental comfort, and their concentration in class \cite{fernandes2011does}. For example, a spacious seat offers physical comfort, a seat near the door results in many distractions, and a seat close to the teacher can result in more attention from the teacher \cite{fernandes2011does, grimm2020teacher}.  
In addition, peers have an impact on student engagement, most especially a negative impact, as peers may invoke more non-academic interactions and off-task behaviours \cite{fernandes2011does, 2019JoshiMultimedia}. However, peers can also have a positive impact by encouraging active learning through discussion \cite{fernandes2011does, bandura1977social}. In general, all behaviours or influences related to the seating experience affect student engagement in class, an essential psychological representation. \textit{Student engagement} refers to the degree of student involvement or interest in learning, as well as their degree of connection with the class or each other \cite{axelson2010defining}. The study of student engagement has attracted growing interest to address problems, such as high dropout rates and declining motivation and achievement \cite{fredricks2004school,fredricks2012measurement}. Specifically, Fredricks et al. \cite{fredricks2004school} identified three dimensions to student engagement: {emotional engagement (i.e. interest, enjoyment and enthusiasm \cite{trowler2010student,fredricks2012measurement}), cognitive engagement (i.e. concentration and comprehension \cite{corno1983role,fredricks2012measurement}) and behavioural engagement (i.e. effort and determination \cite{finn1995disruptive,fredricks2004school, di2018engagement})} .

There is a large body of research investigating the impact of seating location on student engagement and classroom experience \cite{shernoff2017separate, 2019JoshiMultimedia, becker1973college,holliman1986proximity}.
Shrenoff et al. \cite{shernoff2017separate} conducted a study in a large university lecture hall and found that 
students who sit in the back report the lowest perceived engagement, and students who sit in the front report the highest  engagement. Joshi et al. \cite{2019JoshiMultimedia} analyzed the influence of multimedia and seating location on academic engagement, and indicated that students who sit close to the multimedia screen pay more attention than students in the middle row. Both  Holliman et al. \cite{holliman1986proximity} and Becker et al. \cite{becker1973college} found that student performance declined as teacher-to-student seating distance increased. 

However, one common limitation of previous studies is that they relied on the self-report survey or \textit{Ecological Momentary Assessment} (EMA) \cite{shiffman2008ecological} as an engagement measurement, which may prone to subjectivity and various response biases (e.g. social desirability and extreme rating bias) \cite{gao2021investigating}. By investigating the accuracy of self-report information, Moller et al. \cite{moller2013investigating} pointed out that the self reports should not be trusted blindly and researchers must take into consideration that the responses can be unreliable. To overcome this limitation, sensing physiological signals (e.g. electrodermal activity) could be an alternative or even a better method for understanding the relationship between seating location and student engagement. {Electrodermal activity (EDA), also known as \textit{Galvanic Skin Response} (GSR), has been used in physiological and psychophysiological research since the 1880s \cite{boucsein2012electrodermal}. It refers to changes in the electrical conductance of the skin in response to sweat secretion, which is controlled by the \textit{Sympathetic Nervous System} (SNS)  \cite{Boucsein2012publication}. The primary process of the SNS is to stimulate the body's \textit{fight or flight responses} \cite{brodal2004central}. When the sympathetic nervous system is highly aroused, the activity of the sweat glands increases, which in turn increases skin conductance \cite{carlson2007physiology}. }

In this way, EDA is widely used to measure arousal in psychology, which is a broad term representing overall activation and is recognised as one of the two dimensions (i.e. arousal and valence) of emotion responses \cite{russell1980circumplex}. Although measuring arousal is not exactly the same as measuring emotion, it is an important component of emotion, and it has been found to be a strong predictor of attention, perception and cognitive processing \cite{storbeck2008affective, turkileri2021emotional}. 
As an indirect method for measuring arousal and increased mental workload, EDA has been adopted to evaluate engagement in various domains, e.g. emotional engagement \cite{gashi2019using, di2018engagement}, student engagement \cite{gao2020n}, game engagement \cite{huynh2018engagemon}, social engagement \cite{hernandez2014using}. However, most researchers studied engagement on the individual level using either behavioural or physiological patterns. To our knowledge, no prior research has explored engagement at the group level or investigated distinct clusters of physiological signals from users who share similar engagement levels.  

On the other hand, in order to facilitate the statistics of the spatial data (seating locations) in questionnaires, most research focused on the general locations (e.g. front, middle, back) in traditional seating arrangement types (e.g. grouped tables or rows-and-columns) rather than the exact seating locations in flexible seating arrangements scenarios \cite{meeks2013impact,holliman1986proximity,shernoff2017separate,2019JoshiMultimedia}. {Furthermore, while some research has been carried out on individual seating experiences, only a few studies have investigated the social aspects of these seating experiences. To deal with above issues, in this research,  we collect the accurate seating location of students with flexible seating arrangements. In addition to investigating individual experiences, the seating experience and engagement are also investigated in a group-wise manner, i.e. student peers and student groups. }  

Therefore, in this research, we aim to answer the following research questions: {\textit{1. How does seating proximity between students relate to their perceived learning engagement? 2. How do students' group seating behaviours relate to their physiologically measured engagement level (i.e. physiological arousal and synchrony)?} } This research contributes to empirical evidence on how the classroom seating experience affects student engagement. In this study, seating experience is defined as seating location (e.g. in the front or back of the classroom). We present the results of a field study in a high school in which we collected survey and wearable data from 23 student participants attending 10 different courses over four weeks. We investigate the relationship between the student seating experience and student engagement.{ The results show that students who sit close together are more likely to have similar engagement levels than those who sit far apart. In summary, the contributions of this research are as follows:}

\begin{itemize}
    \item We investigate how student seating experience affects student engagement by understanding their perceived engagement {and physiologically measured engagement, measured by EDA signals}. A field study was conducted on a high school campus, with 23 student participants attending 10 courses over hundreds of classes. During the four-week data collection, each participant was asked to wear the E4 wristband during school and report their learning engagement and seating location in the classroom.
    
    \item  To the best of our knowledge, we are the first to study how individual and group-wise classroom seating experiences relate to student engagement. We analyse student engagement from two different perspectives: perceived engagement and physiologically measured engagement.
    
    \item  For the first time, we identify statistically significant correlations between student seating behaviours and students' perceived and physiologically measured engagement. {Our results show that students who sit close together  are more likely to have similar learning engagement and physiological synchrony than students who sit far apart.} 

\end{itemize}

The remainder of the paper is as follows. Section \ref{sec:relatedworks} introduces related works of student seating choices and learning engagement in educational literature, and recent progress of understanding student engagement using physiological signals. Section \ref{sec:datacollection} describes the data collection procedure and measures, and the description of collected seating behaviour data and physiological signals. Then, we show the overall distribution of student seating behaviours, learning engagement,  and physiological patterns in Section \ref{sec:overview}. In Section \ref{sec:relationship}, we explore the relationship between student seating behaviours and their physiological-measured engagement. 
In Section \ref{sec:discussion}, we discuss the implications and limitations of this research. Finally, we summarize the research in Section \ref{sec:conclusion} and point out the potential directions in future studies.

\section{Related Work}

This section described related works on student engagement and the classroom seating experience in educational settings. A brief summary of the main related works could be found in Table ~\ref{tab:relatedwords}.
\label{sec:relatedworks}
\subsection{Student Engagement in Educational Research}

The concept of student engagement has a history from ten to seventy years \cite{axelson2010defining}. In the 1930s, the educational psychologist Ralph Tyler began to exploring the time students spent at work and its impact on learning. Harper et al. \cite{harper2009institutional} argued that engagement is more than participation or involvement; it requires feelings, sense-making as well as activities. In 2004, drawing on Bloom's research \cite{bloom1956taxonomy}, Fredricks et al. \cite{fredricks2004school} identified three dimensions of student engagement: { (1) \textit{emotional engagement}. Students who are emotionally engaged would experience affective reactions, such as interest, enjoyment or a sense of belonging \cite{trowler2010student,fredricks2012measurement}; (2) \textit{behavioural engagement}. Behaviourally engaged students usually abide by behavioural norms, such as attendance and participation, and exhibit an absence of destructive or negative behaviour \cite{finn1995disruptive,fredricks2004school}; (3) \textit{cognitive engagement}. Students who engage cognitively would be invested in their learning, show a willingness to go beyond the requirements and exert efforts to comprehend complex ideas \cite{corno1983role,fredricks2012measurement}.} 

{Student engagement at a particular school or university is increasingly recognised as an effective indicator of institutional excellence, rather than traditional characteristics (e.g. the numbers of books in the library or number of Nobel laureates in the faculty). The self-report survey is one of the most widely used tools for measuring student engagement, and some of the most popular are the \textit{National Survey of Student Engagement} (NSSE) \cite{axelson2010defining}, \textit{Engagement vs. Disaffection with Learning} (EvsD) \cite{skinner2009motivational}, 
\textit{Motivated Strategies for Learning Questionnaire (MSLQ)} \cite{pintrich1990motivational}, and \textit{School Engagement Measure} (SEM) \cite{moore2005conceptualizing}. Recently, some short questionnaires (e.g. \textit{In-class Student Engagement Questionnaires} (ISEQ) \cite{fuller2018development} ) have been designed for \textit{Experience Sampling} \cite{larson2014experience}, which reduces the problem of recall failure and provides instant feedback to inform a cycle of quality improvement.}

\subsection{Classroom Seating Experience and Student Engagement}
\label{sec:related_seating}
\subsubsection{Flexible Seating in Classroom}
Standard classrooms are set up in the traditional linear seating arrangement, {with standard desks and chairs facing the podium.}  However, pedagogical studies \cite{fernandes2011does, burgeson2017flexible, montello1988classroom, 2018MsT19Aflexibles, 2019JoshiMultimedia, grimm2020teacher} have found that when compared with traditional seating, flexible seating provides a more comfortable environment and has {multiple benefits that improve educational activities}. Flexible seating uses various seating options \cite{grimm2020teacher} (e.g.  balls or cushions and standing or seating options) or non-linear seating arrangements (e.g. u-shaped or semicirclular) \cite{fernandes2011does,2019JoshiMultimedia}. Yang et al.\cite{Yang2021Arrangements} found that in English learning courses, the semicircular arrangement facilitated student engagement by enhancing communication, concentration and the classroom environment{ when compared with the traditional arrangement}. In addition, researchers \cite{fernandes2011does, ngware2013influence, 2021academicdisciplineKa} have suggested that tailoring classroom seating arrangements to educational activities helps manage students and results in better perceptions of student behaviours.
\begin{table}[]
\footnotesize
\caption{Related works that studied student seating experience and learning engagement, performance and emotion.}
\label{tab:relatedwords}
\begin{tabular} { p{0.6cm} p{2.2cm} p{2.2cm} p{3.1cm} p{2.3cm} p{3cm} }
\toprule
\textit{Ref.}               & \textit{Seating Options}                          & \textit{Seating Data}                           & \textit{Assessed Item}                                              & \textit{Participants}                               & \textit{Data Type}                                                    \\ \midrule
\cite{2021academicdisciplineKa}        & Rows-and-columns, self-select               & Accurate x, y position in the classroom & Academic performance, engagement                            & 182 university students from 4 disciplines  & Academic score                                               \\\hline
\cite{2019JoshiMultimedia} & Rows-and-columns, self-select               & Rows and groups                        & Academic performance, attention                             & 25,000+ university students       & Attendance, academic score, head-down activity, eye activity \\\hline
\cite{shernoff2017separate}          & Rows-and-columns, self-select, semi-circular & Front, middle, back of the classroom   & Class experience, engagement, course performance             & 407 university students                    & self-reported engagement, attention, classroom experience, course grade    \\\hline
\cite{2020GaiannakosFitbit}         & N/A                                      & N/A                                    & Academic performance, study satisfaction, study usefulness & 31 university students (93 sessions)                     & Self-reported experience, HR, EDA, TEMP, BVP                 \\\hline
\cite{gashi2018using}             & N/A                                      & N/A                                    & Class experience, emotion state                             & 24 university students (1008 sessions)                   & Self-reported experience and emotional state, EDA            \\\hline
\cite{di2018engagement}            & N/A                                      & N/A                                    & Emotional engagement                                       & 24 university students (984 sessions)                    & Self-reported emotional engagement, EDA, BVP                 \\
 \bottomrule
\end{tabular}
\end{table}

\subsubsection{Seating Preference and Student Engagement}

Most researchers concluded that seating location has an effect on student engagement, attention, involvement and motivation \cite{montello1988classroom,ngware2013influence,burda1996college}. Ngware et al. \cite{ngware2013influence} showed that students' seating locations were related to their academic abilities. Sitting in the front row led to greater learning gains (5\%-27\%) when compared with sitting far away from the front. Burda et al. \cite{burda1996college} indicated that students who choose the back seats may be more passive and feel more comfortable sitting far away from the teacher to ensure less interaction. These students were  often observed disengaging from the class. Gyanendra et al. \cite{2019JoshiMultimedia} found that university students who preferred to sit in the first or the last two rows of the classrooms paid greater attention to the multimedia screen and had higher grades than those who sat in the middle. 

{Kalinowski \cite{kalinowski2007effect} compared academic scores with students' seating preferences and assigned seating locations.} They found that students with higher GPAs preferred to sit in the front. This suggests that a correlation between seating location and motivation. Ka et al. \cite{2021academicdisciplineKa} examined the relationship between the seating location and academic performance of 182 university students from four disciplines. It highlighted a significant relationship between seating location and academic performance. {However, the relationship varied between the academic disciplines. In particular, in fields requiring more active and integrated learning, such as nursing, students who sat in the front performed batter academically than those who sat at the back because of their higher level of participation and engagement.} It is worth noting that the only the academic scores were compared, which is not sufficient to represent academic performance or engagement.

\subsubsection{Seating Proximity and Student Engagement}
{The proximity of the student to the teacher \cite{millard1980enjoyment,shernoff2017separate} and student groups \cite{dong2021influence,gremmen2018importance} can affect student engagement and satisfaction in the classroom. Millard et al. \cite{millard1980enjoyment} found that when undergraduates were periodically moved from one location to another, students' enjoyment and productivity changed significantly in both free and assigned seating settings. They also found that increasing proximity of student and teacher was related to decreasing self-reported motivation, enjoyment, interests and feeling apart.
Shernoff et al. \cite{shernoff2017separate} suggested that seating location and distance from the teacher are consistently correlated with student engagement, attention and course performance. 
Although social interactions, such as group discussions, positively impact engagement and encourage active learning \cite{hurst2013impact}, there is a risk that non-academic interactions also increase, thereby distracting students. 
Teachers can assign seats in an attempt to control student interactions, e.g. specifically assigning seats to reduce non-academic peer interactions, which have a negative influence on academic achievement \cite{fernandes2011does}. Gremmen et al. \cite{gremmen2018importance} investigated whether near-seated peers influence students' academic engagement and achievement in elementary school settings. They found that students achieved better (worse) scores when near-seated friends scored better (worse).}

{Social learning theory states that people learn by observing and imitating others \cite{bandura1977social}. Based on this theory, students should learn by observing peers \cite{bandura1977social}, and peer conditions affect student engagement and motivation. Gyanendra et al. \cite{2019JoshiMultimedia} indicated that students preferred to choose seats with similar proximity to the multimedia screen, and the students who sat at a similar proximity to the front had similar distraction rates and performance levels \cite{2019JoshiMultimedia}, e.g. students with higher grades chose to sit in the front rows. These findings suggest a bidirectional relationship, i.e. performance (or motivations) connects peers, and peer conditions impact performance (or motivation). 
Although arguably, grades are correlated with seating location \cite{2019JoshiMultimedia, ngware2013influence, montello1988classroom, 2021academicdisciplineKa}, it was empirically discovered that study performance and engagement significantly improved when students sat closer \cite{fernandes2011does, gashi2019using}. Netware et al. \cite{ngware2013influence} also noted  the correlation between seat location and in-class engagement, and suggested that teachers could optimise their teaching effectiveness by monitoring the progress of students sitting in different rows. }


\subsection{Inferring Student Engagement Using Sensing Technologies}
\label{sec:related_sensing}

Recently, physiological signals, e.g. EDA \cite{gao2020n,di2018engagement,wang2015physiological}, PPG \cite{monkaresi2016automated,gao2020n},  electroencephalogram (EEG) \cite{szafir2013artful}, eye gaze \cite{d2012gaze} and facial expressions \cite{woolf2009affect}, have been used to infer learners' affective states. {Especially, there is a thread of research using EDA to indicate engagement levels \cite{gashi2019using, di2018engagement,gao2020n,huynh2018engagemon,hernandez2014using,ward2018sensing,nan2022human}. Hernandez et al. \cite{hernandez2014using} measured the engagement of the child during interaction using physiological synchrony extracted from EDA sensors. Lascio et al. \cite{di2018engagement} measured university students' emotional engagement from EDA signals. Huynh et al. \cite{huynh2018engagemon} developed EngageMon  to use EDA together with HRV, touch and vision to indicate game engagement. Gao et al. \cite{gao2020n} predicted 
student multi-dimensional engagement using EDA, PPG and HRV signals.}


Physiological synchrony indicates the observed association or interdependence between the physiological processes of two or more people. These physiological signals often reflect connections between people's continuous measurements of the autonomic nervous system \cite{palumbo2017interpersonal}. Across various streams of research, physiological synchrony has been shown to be informative for cognitive demands, task difficulties, learning engagement, etc. {Therefore, understanding the physiological synchrony between people has attracted attention in the ubicomp community \cite{gashi2018using, ward2018sensing, malmberg2019we, stuldreher2020multimodal}. }
Gashi et al. \cite{gashi2018using} proposed using physiological synchrony between students to measure the classroom emotional climate (CEC). They calculated the group physiological synchrony by applying the Dynamic Time Warping (DTW) distance to processed EDA signals and found that the group physiological synchrony of EDA was positively correlated to the CEC. Gashi et al. \cite{gashi2019using} studied the physiological synchrony of the inter-beat interval (IBI) and EDA signals between a presenter and the audience. They found that these signals could be used as a proxy to quantify participants' agreement on self-reported engagement during presentations. Malmberg et al. \cite{malmberg2019we} found that physiological synchrony occurred within two groups of students who experienced difficulties in collaborative exams and concluded that physiological synchrony may be an indicator of the recognition of meaningful events in computer-supported collaborative learning.

{
In summary, unlike previous studies, our research has the following advantages: (1) We are the first to examine how student seating behaviour is related to physiologically measured student engagement rather than simply using traditional measures (i.e. perceived student engagement) as previous studies have done \cite{shernoff2017separate,2019JoshiMultimedia,sanders2013engagement,2020GaiannakosFitbit}. (2)
We explore student seating behaviours using exact seating locations, which provides greater flexibility, whereas most studies only explore traditional seating arrangement styles (e.g. grouped tables or traditional row arrangement) \cite{shernoff2017separate,2019JoshiMultimedia,sanders2013engagement}.  (3) We explore the student engagement at the group level by utilizing distinct clusters of physiological signals, while previous researchers have studied student engagement at the individual level \cite{di2018engagement,gao2020n}. (4) For the first time, we identify some statistically significant correlations between student seating behaviours and students' perceived and physiologically measured engagement.}

\section{Data Collection}
\label{sec:datacollection}

\subsection{Study}
We collected data from a field study \cite{gao2022understanding} in a high school over four weeks. The study has been approved by the Human Research Ethics Committee at the researchers' institutions, which was furthermore approved by the principal of the high school. We have recruited 23 students (15-17 years old, 13 female and 10 male) and 6 teachers (33-62 years old, 4 female and 2 male) in Year 10. After returning the signed consent forms by teachers and students (and their guardians), the participants were asked to complete a {background survey to report their demographic and class information. With the class information provided by students and class schedule provided by the school, we were able to locate the student in specific classrooms in any time period. }

\begin{figure}
    \centering
    \includegraphics[width=.18\textwidth]{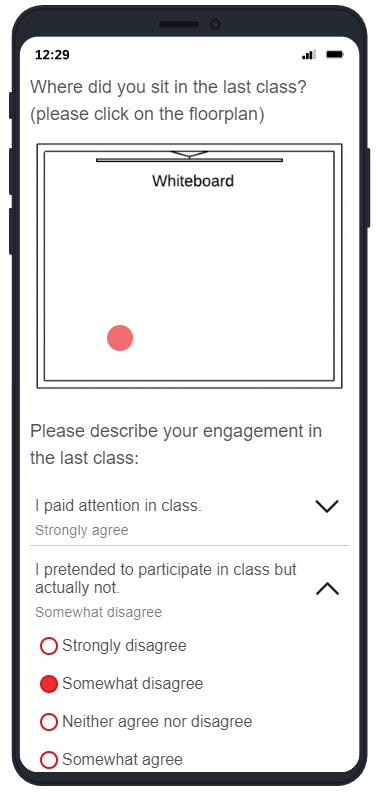}
    \caption{The screenshot of the self-report survey.}
    \label{fig:screenshot}
\end{figure}

Before the data collection, all \textit{Empatica E4} \footnote{Empatica E4: \url{https://www.empatica.com/en-int/research/e4/}} wristbands were synchronized with the E4 Manager App from the same laptop to make sure the internal clocks were correct. {There are two modes for E4 wristbands to collect data: recording mode and Bluetooth streaming model. We used the recording mode where the raw data were saved in the internal memory which allows recording for up to 60 hours.
In a word, E4 wristband has a long standby time (more than 32 hours), large storage space, is lightweight, waterproof and comfortable to wear, making it suitable for continuously unobtrusive monitoring in this research. The reason that we did not use consumer-oriented wearable devices is that most of them use undisclosed algorithms and only provide summaries or proprietary metrics rather than raw data, which brings much uncertainty for scientific research. }

{To protect the personal privacy, we anonymised students' information by assigning each student an ID. }During the data collection period, student participants were asked to wear the same E4 wristband (attached with a sticker with the student ID) on the non-dominating hands at school time. {They were told to avoid pressing buttons or performing any unnecessary movements during class.} Additionally, they were reminded by the class representative to complete online surveys three times a day at 11:00, 13:25, 15:35 (right after the second, fourth and fifth class). Teacher participants only need to wear the wristband during their classes and complete an online survey right after their class. {Overall, we have collected 488 survey responses and 1415 hours of wearable data from all participants}. As a token of appreciation, participants have distributed four movie vouchers for 4-week data collection. Participation in this research project was completely voluntary, and participants were free to withdraw from the project at any stage.

\subsection{Measures}
{We collected student seating locations and multi-dimensional engagement from self-report surveys completed on public tablets or participants' own smart devices, using the link generated by \textit{Qualtrics}. We also collected  physiological signals using E4 wristbands during the school time.}

\subsubsection{Student Engagement} We used a self-report survey to collect subjective assessments of student engagement. The self-report tool is the most commonly used method to measure student engagement {because it can clearly reflect subjective perceptions, while other methods, such as interviews, teacher ratings and observations, are susceptible to external factors \cite{fredricks2004school,fredricks2012measurement}}. The student engagement questionnaire includes five items from the validated ISEQ questionnaire \footnote{{The original questions are: (1) I devoted my full attention; (2) I pretended to participate; (3) I enjoyed learning new things; (4) I feel discouraged; (5) The activities really helped my learning; (6) I tried a new approach or way of thinking about content.} }\cite{fuller2018development}, which was proved to be effective for multidimensional engagement measurement compared with the traditional long survey. Similar to previous research \cite{huynh2018engagemon,di2018engagement}, we slightly adapted the survey questions \footnote{{Specifically, the modified questions are: (1) I paid attention in class; (2) I pretended to participate in class but actually not; (3) I enjoyed learning new things in class; (4) I felt discouraged when we worked on something; (5) I asked myself questions to make sure I understood the class content. Question 1,3 and 5 assess the behavioural, emotional and cognitive engagement respectively, where item 2 and 4 indicate the
behavioural and emotional disaffection   \cite{fuller2018development,skinner2009motivational}.}} to suit high school classes and make it easier for underage students to understand. In addition, we did not adopt the original questions ‘\textit{The activities really helped my learning of this topic}’ {and ‘\textit{I tried a new approach or way of thinking about content}’} for assessing cognitive engagement in \cite{fuller2018development}, because some high school classes may not always teach new content or have in-class activities. Instead, we used the well-accepted question ‘\textit{I asked myself questions to make sure I understood the class content}’ \cite{moore2006children}, which has been proven to be a good reflection of cognitive engagement.
In the questionnaire, each item was rated on a 5-point Likert-scale from ‘strongly disagree’ to ‘strongly agree’.

\begin{figure}
    \centering
    \includegraphics[width=0.92\textwidth]{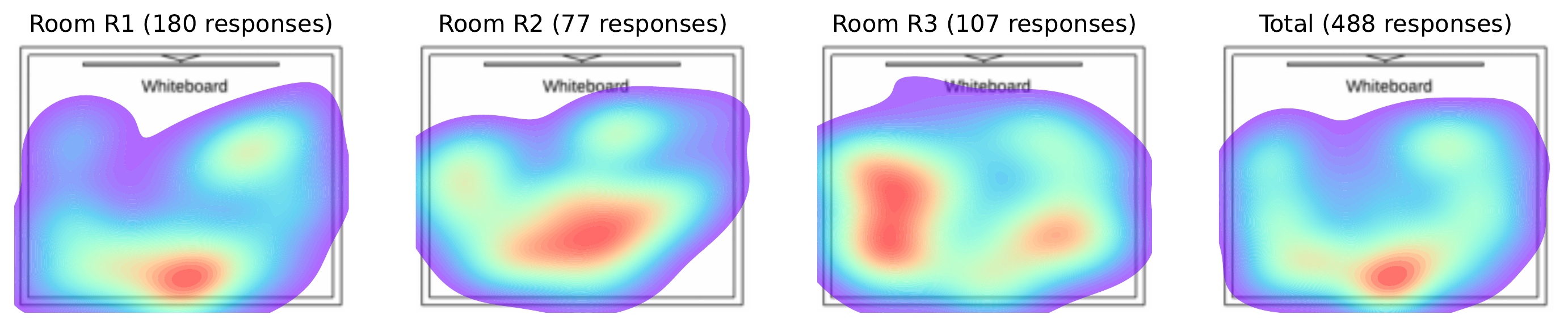}
    \caption{Overall seating distribution of students in the classrooms.}
    \label{fig:seating_room}
\end{figure}
\subsubsection{Seating Location}
{Classrooms for students in Year 10 are approximately 7.0 m $\times$ 8.9 m in size and can accommodate up to 25 students. This school encourages flexible seating arrangements, therefore seating arrangements will often vary based on teacher needs, teaching style and course content. For example, rows and columns are adopted when individual works are preferred, small groups are organised when more interaction is encouraged, semi-circular or u-shape seating arrangements are used when communication is emphasized. Before each class, students are free to choose where they want to sit in the classroom, leaving multiple vacant seats as some students may be absent for various reasons}. During the data collection, the seating location of participants was measured by the self-report item '\textit{Where did you sit in the last class? (Please click on the floor plan)}'. Participants were shown floor plan pictures \footnote{All classrooms in Year 10 have  the same floor plan.} and they could click different locations to report where they sat (see Figure \ref{fig:screenshot}). The seating location was recorded as the $x$ and $y$-coordinate (in pixels) for each click, where $x = 0, y = 0$ represented the upper left corner of the floor plan. {Compared to traditional methods of asking students about general locations (e.g. back/middle/front of the room \cite{shernoff2017separate}, districts of multiple rows \cite{2019JoshiMultimedia}) or the sequential number of rows and columns  \cite{ngware2013influence,lyu2021relations}, reporting the exact location in a classroom enables us to understand seating behaviours in real-world scenarios with flexible seating arrangements.} Figure ~\ref{fig:seating_room} shows the heat map of the seating locations from different classrooms in Year 10.

\subsubsection{Physiological Signals}
We assessed participants' EDA signals using the \textit{Empatica E4} wristbands. {The E4 wristband is equipped with multiple sensors designed to gather high-quality data and has an EDA sensor, PPG sensor, an ACC and an optical thermometer}. EDA sensors record the constantly fluctuating changes in the electrical properties of the skin at 4 Hz. When the level of sweat increases, the conductivity of the skin increases. For most people, when they experience increased cognitive workload, emotional arousal or physical exertion, the brain will send innervating signals to the skin to increase the sweat production. Therefore, even though they may not feel any sweat on the skin surface, conductivity increases noticeably.
Specifically, EDA complex includes two main components: a general tonic component to measure the skin conductance level (SCL) and rapid phasic component to measure the skin conductance response (SCR) resulting from sympathetic neuron activity \cite{guideeda}. The SCL measures the slow-acting and background characteristics of the EDA signal (overall level and slow decline or increase over time), reflecting the influence of autonomic arousal on the general sweat glands. The SCR is usually a sudden increase in the skin's conductance, which is usually associated with short-term events and external/internal stimuli.

\begin{figure}
    \centering
    \includegraphics[width=0.99\textwidth]{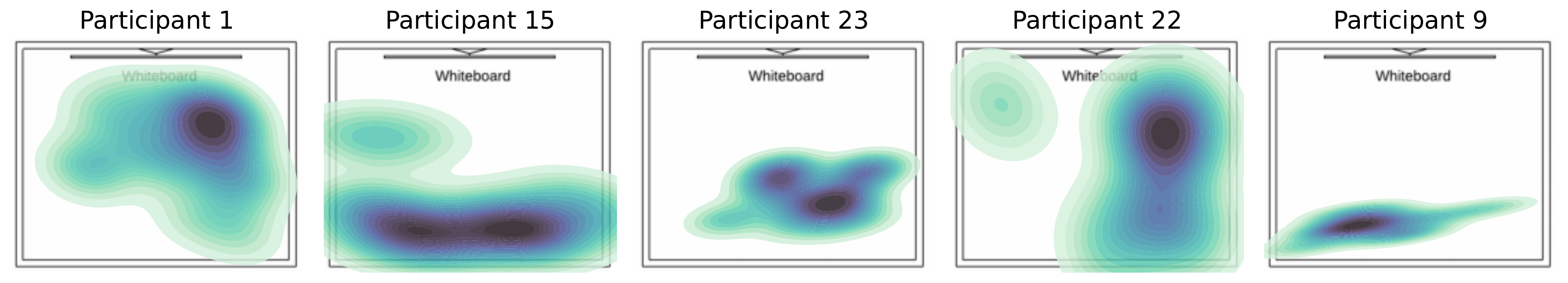}
    \caption{Seating location across five different student participants.}
    \label{fig:loc_dis}
\end{figure}

\section{Overview of Seating Experience, Student Engagement, and Physiological Patterns}
\label{sec:overview}
{In this section, we explored the patterns of seating experience, student engagement and physiological signals. We divided the student groups based on seating locations and investigate the seating preference of participants over different courses. Then, we displayed the distribution of student engagement and the number of collected wearable signals across all courses.}

\subsection{Seating Locations and Seating Preference}
Figure \ref{fig:loc_dis} displays the seating locations of five different students participants: P1, P15, P23, P22 and P9. Due to space limitations, we have not shown the seating distribution of all participants. We found that different participants tended to have very different seating preferences. For instance, participant P22 typically sat close to the whiteboard, while participant P9 tended to sit at the back of the classroom. {To better investigate the seating locations of students, we divided the students' seats into different groups. The simplest and most intuitive ways to do this is to partition the classroom equally from left to right or front to back. For example, sitting on the left-hand side of the classroom centre line is regarded as the left. }

\begin{figure}[b]
    \centering
    \includegraphics[width=0.8\textwidth]{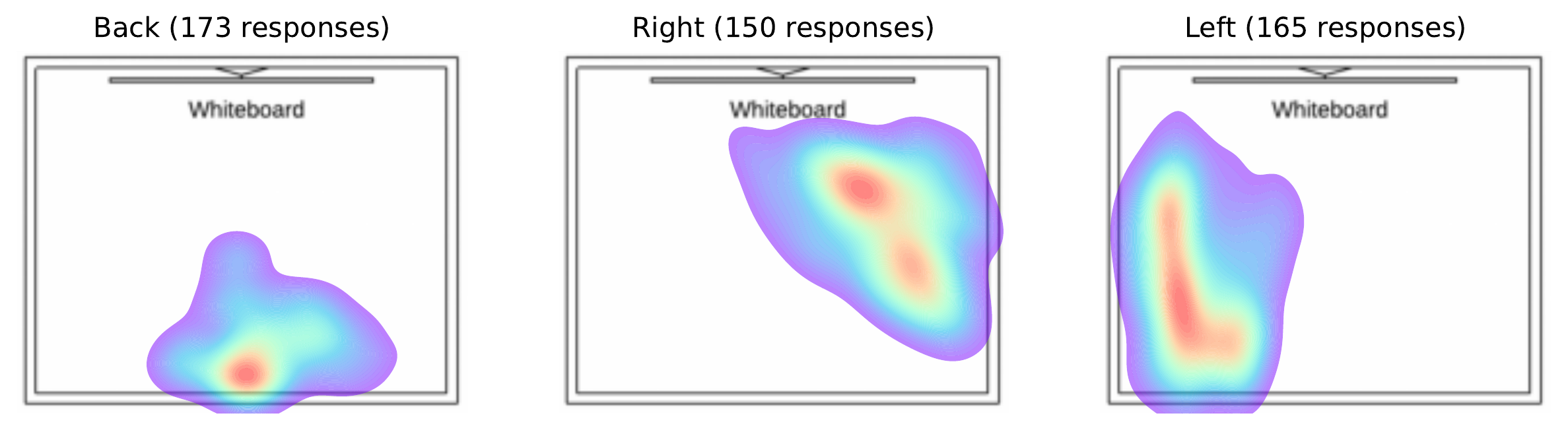}
    \caption{Three different seating locations (back, right and left) in classrooms.}
    \label{fig:cluster_pos}
\end{figure}

{However, this simple partitioning method is not applicable in this research for two reasons: 1) The seats in the classroom were not evenly distributed from left to right. For example, the seats were sometimes arranged in two semicircles, with one semicircle positioned towards the center of the classroom and the other towards the side of the classroom. 2) The seats in the classroom were not evenly distributed from front to back. First, the whiteboard occupies a large area in the front of the classroom, which means dividing the classroom evenly from front to back may result in a greater concentration of chairs in the back half of the classroom. More importantly, the seats were not always arranged symmetrically like in traditional seating arrangements (e.g. row-and-column and grouped tables), and many times, they were distributed in a u-shape configuration according to the needs of teacher.}

{Therefore, we chose to group seating locations using the clustering technique. As one of the most popular clustering methods, we employed the \textit{k-means} algorithm \cite{kmeans} to group similar seating locations and discover underlying patterns.} However, for the \textit{k-means} algorithm to be effective, the number of the clusters must be predefined. There are two popular ways of determining the optimal number of clusters: the Elbow method and the Silhouette method \cite{optimalk}. 
After calculating the number of clusters using both methods, we identified that the optimal number of clusters was three. We then ran the \textit{k-means} algorithm using the \textit{Scikit\-learn} \cite{pedregosa2011scikit} Python package with \textit{n\_clusters} = 3 and \textit{random\_state} = 0. Figure~\ref{fig:cluster_pos} shows the clustering results of the seating locations from 488 responses. From these diagrams, it is clear that there were three different seating preferences in the classes: back, right and left. Therefore, in this research, we mainly focus on those three seating preferences.

\begin{figure}
    \centering
    \includegraphics[width=0.68\textwidth]{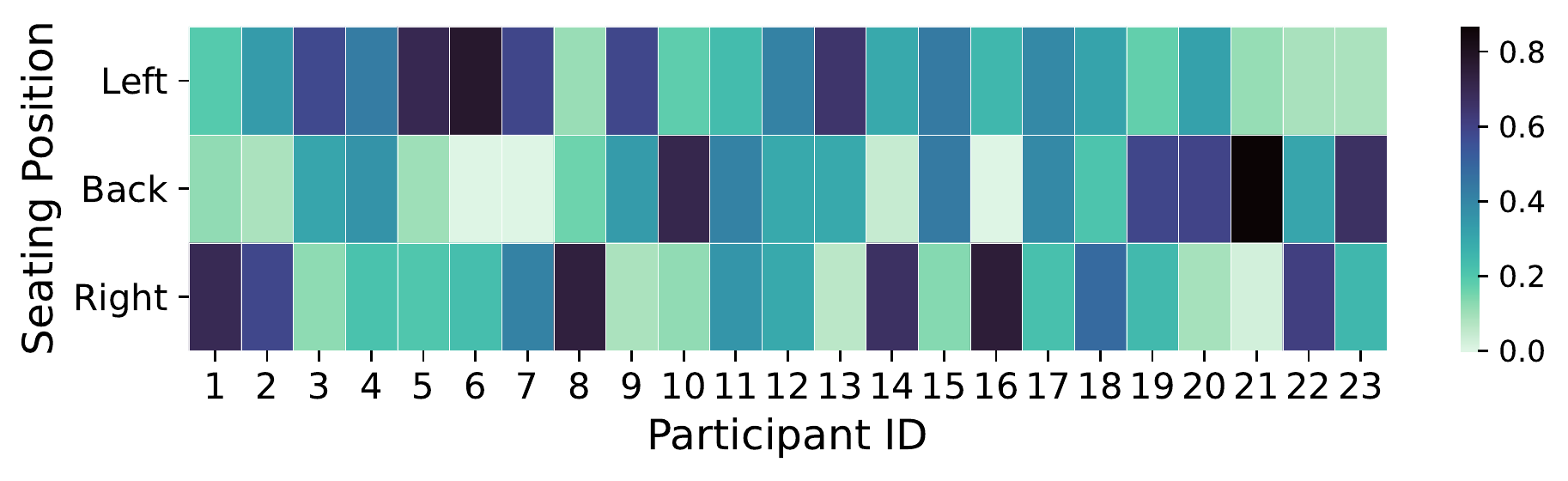}
    \caption{Seating preference for each participant in all courses.}
    \label{fig:seating_per}
\end{figure}
\begin{figure}[b]
    \centering
    \includegraphics[width=1\textwidth]{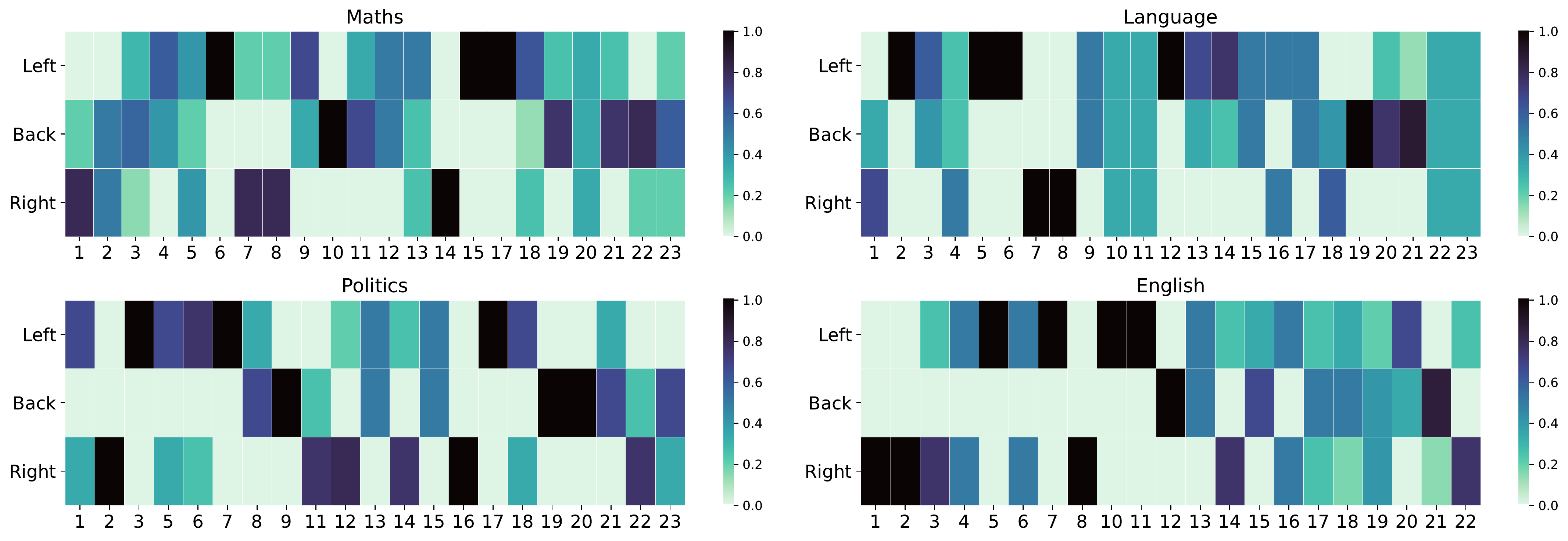}
    \caption{Seating preference for each participant in each course.}
    \label{fig:seating_per_courses}
\end{figure}

An overview of the seating preferences of each participant is shown in Figure ~\ref{fig:seating_per}, and different colours indicate the frequency of sitting in each area. It shows that different participant tended to have very different seating preferences. For example, some participants (e.g. P10 and P21) usually sat in the back of the classroom, while some (e.g. P6 and P16) sat the left or right. Interestingly, some participants (e.g. P11 and P12) did not have obvious preferences for where they sat.
Figure \ref{fig:seating_per_courses} shows how the seating preferences of the participants varied by course. We found that participants often had different seating preferences in different courses, e.g. P3 never sat in the back in Politics or English but usually sat in the back in Maths. According to the education research, the potential reasons may be different student motivation and interests \cite{benedict2004seating}, territoriality and the desire to feel comfortable in different learning environments \cite{kaya2007territoriality}, and peer conditions within the classrooms \cite{weaver2005classroom}. 

\subsection{Student Engagement and Physiological Signals}
\begin{figure}
	\centering
	\subfigure[Engagement score across  participants\label{fig:dis_engage}]{\includegraphics[width=0.48\textwidth]{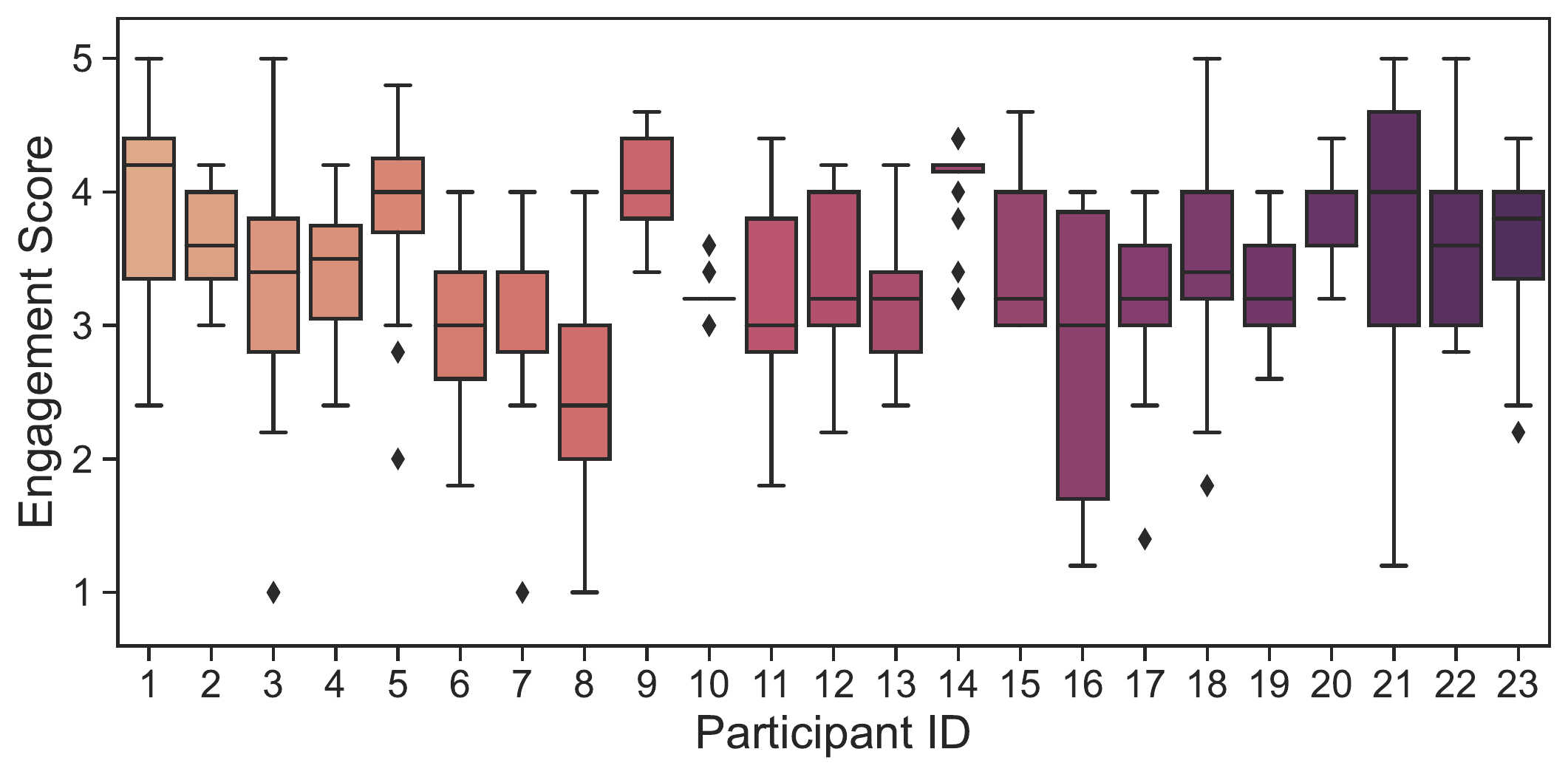}}
	\hspace{0.1cm}
	\subfigure[Engagement score across courses \label{fig:dis_engage_course}]{\includegraphics[width=0.48\textwidth]{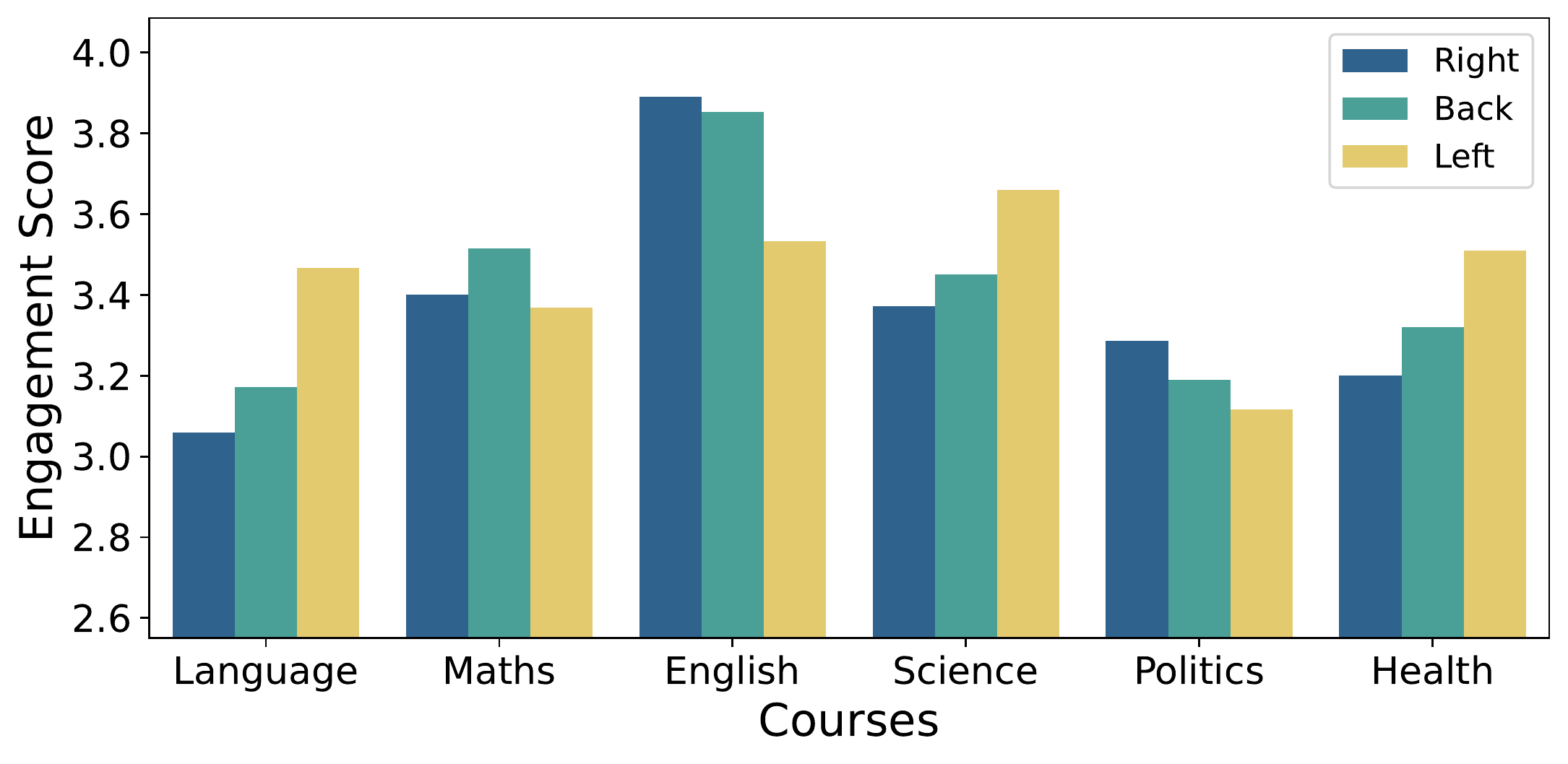}}
    \caption{The overview of engagement score across participants and courses.}
    \label{fig:dis_boxplots}
\end{figure}

The distribution of overall engagement across student participants is shown in Figure \ref{fig:dis_engage}. The overall engagement scores were calculated based on the five items in the questionnaire, where 1 = lowest engagement and 5 = highest engagement. We can see that different participants usually have different engagement scores. Some participants (e.g. P5 and P9) tended to be highly engaged in class most time while some participants (e.g. P16) have varying levels of engagement in different classes. Based on the three seating preferences, the engagement score were calculated across different courses in Figure \ref{fig:dis_engage_course}. 


Then, we calculated the number of wearable signals across all courses in Table \ref{tab: wearable signals}. There were 10 different courses, and students were divided into three groups: Maths, Language and Form groups. 'All' indicates that all students were in one big group. Most of the classes were held in rooms \textit{R1}, \textit{R2} and \textit{R3}. There was an extra room, \textit{R4} for one language group, \textit{R5} was the room for Science, R6 was for Assembly and R7 was for Chapel. 'Out' indicates the playground, which was used for the physical education course. 
For the wearable data, there are a total of 1,120 session logs of signal traces that can be used to explore the physiological patterns related to student engagement and seating preferences.

\begin{table}[b]
\caption{An overview of the number of wearable signals across all courses.}
\begin{tabular}{@{}
>{\columncolor[HTML]{FFFFFF}}l 
>{\columncolor[HTML]{FFFFFF}}l 
>{\columncolor[HTML]{FFFFFF}}l 
>{\columncolor[HTML]{FFFFFF}}l 
>{\columncolor[HTML]{FFFFFF}}l @{}}
\toprule
{\color[HTML]{000000} \textit{Course}} & {\color[HTML]{000000} \textit{Classtype}} & {\color[HTML]{000000} \textit{Room}}  & {\color[HTML]{000000} \textit{Numbers}} & {\color[HTML]{000000} \textit{Signal Traces}} \\ \midrule
{\color[HTML]{000000} Maths}           & {\color[HTML]{000000} Math groups}        & {\color[HTML]{000000} R1, R2, R3}     & {\color[HTML]{000000} 36}               & {\color[HTML]{000000} 197}                    \\
{\color[HTML]{000000} Language}        & {\color[HTML]{000000} Language groups}    & {\color[HTML]{000000} R1, R2, R3, R4} & {\color[HTML]{000000} 44}               & {\color[HTML]{000000} 181}                    \\
{\color[HTML]{000000} English}         & {\color[HTML]{000000} Form groups}        & {\color[HTML]{000000} R1, R2, R3}     & {\color[HTML]{000000} 31}               & {\color[HTML]{000000} 172}                    \\
{\color[HTML]{000000} Politics}        & {\color[HTML]{000000} Form groups}        & {\color[HTML]{000000} R1, R2, R3}     & {\color[HTML]{000000} 37}               & {\color[HTML]{000000} 190}                    \\
{\color[HTML]{000000} Science}         & {\color[HTML]{000000} Form groups}        & {\color[HTML]{000000} R5}             & {\color[HTML]{000000} 32}               & {\color[HTML]{000000} 160}                    \\
{\color[HTML]{000000} Health}          & {\color[HTML]{000000} Form groups}        & {\color[HTML]{000000} R1, R2, R3}     & {\color[HTML]{000000} 18}               & {\color[HTML]{000000} 64}                     \\
{\color[HTML]{000000} PE}              & {\color[HTML]{000000} Form groups}        & {\color[HTML]{000000} Out}            & {\color[HTML]{000000} 13}               & {\color[HTML]{000000} 64}                     \\
{\color[HTML]{000000} Form}            & {\color[HTML]{000000} Form groups}        & {\color[HTML]{000000} R1, R2, R3}     & {\color[HTML]{000000} 6}                & {\color[HTML]{000000} 29}                     \\
{\color[HTML]{000000} Chapel}          & {\color[HTML]{000000} All}                & {\color[HTML]{000000} R7}             & {\color[HTML]{000000} 2}                & {\color[HTML]{000000} 28}                     \\
{\color[HTML]{000000} Assembly}        & {\color[HTML]{000000} All}                & {\color[HTML]{000000} R6}             & {\color[HTML]{000000} 2}                & {\color[HTML]{000000} 35}                     \\
{\color[HTML]{000000} Total}           & {\color[HTML]{000000} N/A}                & {\color[HTML]{000000} R1 - R7, Out}   & {\color[HTML]{000000} 221}              & {\color[HTML]{000000} 1,120}                   \\ \bottomrule
\end{tabular}
\label{tab: wearable signals}
\end{table}

\section{Result}
\label{sec:relationship}

{In this section, we discussed the extensive experiments that were conducted to understand how individual and group-wise seating experiences affect student engagement. We would answer the first research question '\textit{How does seating proximity between students relate to their perceived learning engagement?}' in Section \ref{sec:corr_percieved}. We would answer the second research question '\textit{How do students' group seating behaviours relate to their physiologically measured engagement level (i.e. physiological arousal and synchrony)?}' in Section \ref{sec:result_2} and Section \ref{sec:result_3}. More specifically, in Section \ref{sec:result_2}, we investigated the correlation between the seating location and physiological synchrony. Meanwhile, we studied the group-wise classroom seating patterns and demonstrated how they affect student engagement in different courses in Section \ref{sec:result_3}.}

\subsection{Relationship Between Seating Behaviours and Perceived Student Engagement}
\label{sec:corr_percieved}

As discussed in Section \ref{sec:related_seating}, seating location has been found to be correlated with student engagement. Although some researchers \cite{millard1980enjoyment,shernoff2017separate} revealed how the seating proximity between the student and teacher influences student engagement, satisfaction and course performance, few studies explored how the proximity of students is correlated with learning engagement. In this research, we define the following two terms: \textit{proximity of students} and \textit{similarity of engagement}. \textit{Proximity of students} is calculated as the seating distance $d_s$ between two students, using the Euclidean distance \cite{euclidean}. Euclidean distance is the most commonly used distance measure, which calculates the straight-line distance between two points on a plane and works well on low-dimensional data. Therefore, the formula to calculate the \textit{proximity of student} is as follows:

\begin{equation}
\label{equ:proximity}
    d_s = \sqrt{ {\left ( x_a - x_b \right )}^2 + {\left ( y_a - y_b \right )}^2}
\end{equation}

\begin{figure}[b]
	\centering
	\subfigure[The number of times pairs of students appeared in the same class \label{fig:matrix_id}]{
	
	\includegraphics[width=0.485\textwidth]{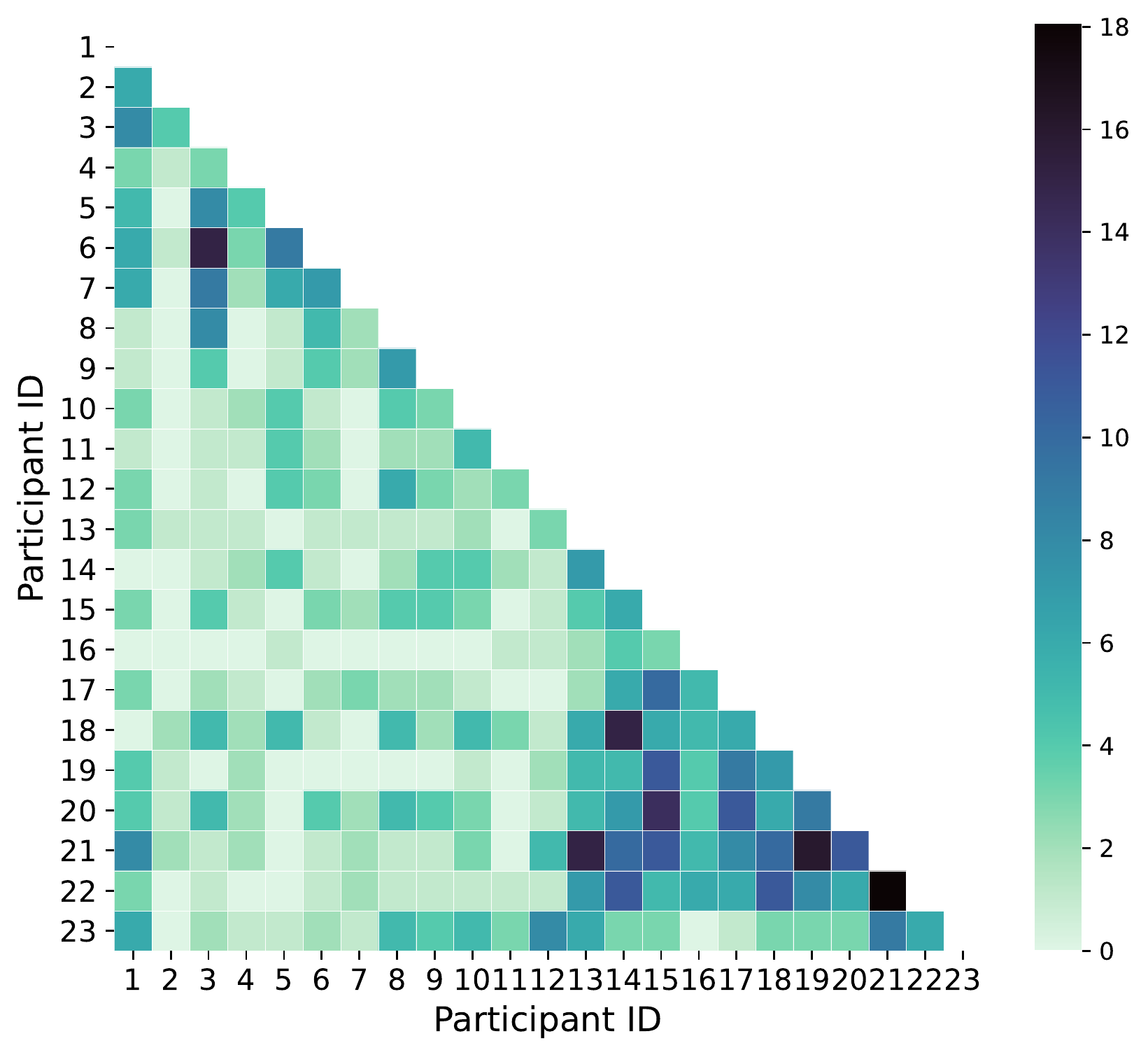}}
	\hspace{0cm}
	\subfigure[The average seating distance between pairs of students\label{fig:matrix_dis}]{
	
	\includegraphics[width=0.485\textwidth]{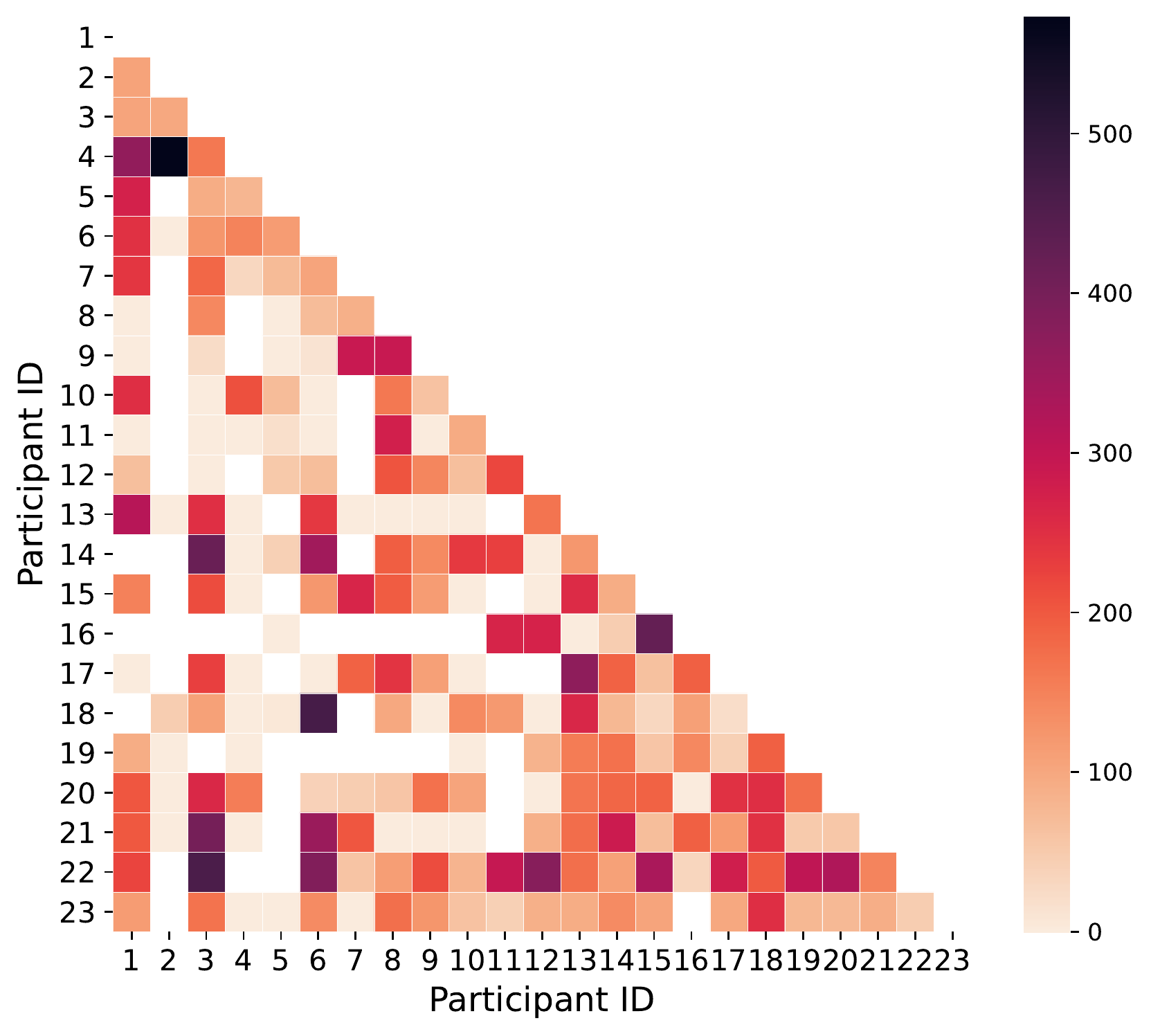}}
    \caption{Seating and occurrence information for pairs of students.}
    \label{fig:matrix}
\end{figure}

In Equation \ref{equ:proximity}, $x_a$ and $y_a$ indicate the position where student $S_a$ sat, {as marked manually by the student, and the upper left point on the figure has the \textit{x}, \textit{y} coordinates (0, 0)}. To measure the \textit{similarity of engagement}, we computed the Manhattan distance \cite{yu2008distance} $d_e$ between the engagement score of two students, $S_a$ and $S_b$, where $d_e = \left | S_a - S_b  \right |, S_a \in [1, 5], S_b \in [1, 5]$. The Manhattan distance works well on discrete/binary variables, and it considers the path that can be realistically taken given the values of the attributes. {Therefore, the \textit{similarity of engagement} $\mathcal{E}(d_e) $ is defined as follows:}

\begin{equation}
\label{equ:sim}
    \mathcal{E}(d_e)  = 1 - \frac{d_e}{4} = 1- \frac{{}|S_a - S_b|}{4}
\end{equation}

In Equation \ref{equ:sim}, $\mathcal{E}(d_e)$ is in the range of [0, 1], where 1 means the participants' engagements were exactly the same, and 0 means the engagements were very different (i.e. one participant is completely engaged while the other one is not engaged at all).

We then analysed the self-report engagement responses and seating behaviours of participants from 92 out of 115 classes (23 classes with responses from only one participant were removed). After removing duplicated responses and keeping the last response from the same participants in each class, we got 1,123 unique pairs of instances (i.e. each instance had a unique combination of two participants' IDs and a class ID). The overall seating and occurrence information for each pair of students is shown in Figure \ref{fig:matrix}. {Figure \ref{fig:matrix_id} shows the number of times the pairs of students appeared in the same class, where a darker colour indicates that the pair usually sat in the same classroom (e.g. P21 and P13). Figure \ref{fig:matrix_dis} shows  average seating distance between pairs of students sitting in the same classroom, where a darker colour means that the two students usually sat far apart. }

\begin{figure}
    \centering
    \includegraphics[width=\textwidth]{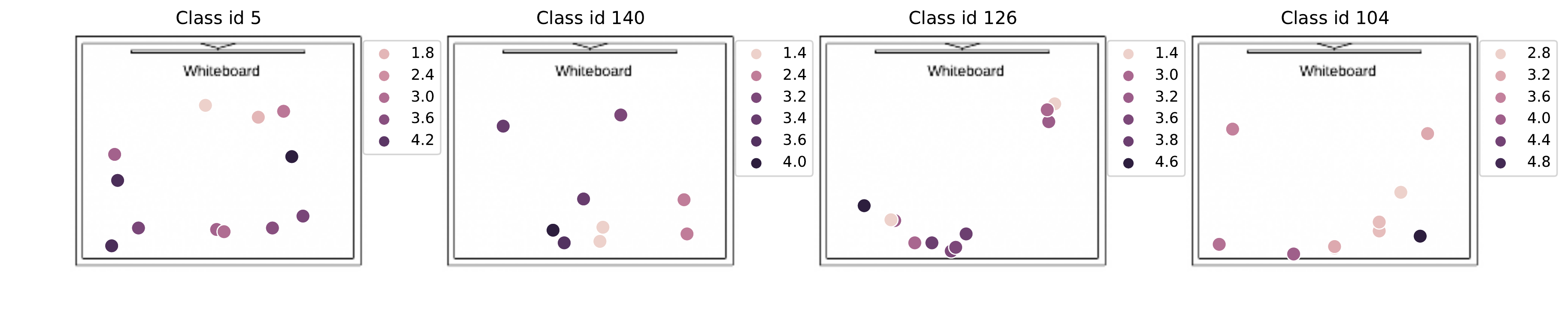}
    \caption{Overall engagement and seating distribution over four example classes.}
    \label{fig:seating_example}
\end{figure}

Figure \ref{fig:seating_example} displays the distribution of the overall engagement and the seating distribution for four example classes. The points indicate the annotated seating locations in the classroom, as marked by different participants. A darker colour indicates a higher overall engagement score, where 1 is the lowest score and 5 is the highest. In particular, in the sub-figure Class ID 140, we observe that points close together have similar colours, i.e. students who sat close together  tended to have similar engagement scores in this class. {It is worth noting that while engagement may change during a class, in this study we only asked students to report their overall engagement for the whole class and complete the survey immediately after the class. Typically, submission times were around scheduled times (i.e. 11:00, 13:25 and 15:35), although they varied slightly for each participant. Therefore, all responses submitted before the next class were considered to be feedback for the previous class.}

Finally, we calculated the correlation between seating proximity and the similarity of engagement for all participants. The experiment result showed that the similarity of engagement was significantly negatively correlated with the seating distance (corr = -0.30 and p < 0.05). This indicates that students who sat close together were more likely to have a similar learning engagement than students who sat far apart.
Interestingly, when considering each individual participant, the similarity of engagement was significantly negatively correlated with the seating distance for 18 out of 23 participants (p < 0.05), moderately negatively correlated with the seating distance for two participants (p < 0.1) and there was no obvious correlation with seating distance for three participants. The possible reason for this result may be due to individual differences, i.e. the engagement of some students may be affected by those around them, while other students may not to be affected by those around them. 

Notably, one potential confounder to the above findings was that the absolute seating position on the whiteboard could affect student engagement \cite{burda1996college, 2019JoshiMultimedia}. Therefore, we calculated the Pearson correlation between the above two variables, where corr = 0.09 (p-value = 0.047), indicating that there is a negligible correlation between student engagement and absolute distance to the whiteboard. A potential reason may be that the seating arrangement is ‘u-shaped’ instead of the traditional ‘row-and-column’ most of the time, students sitting far from the whiteboard usually face the whiteboard, while students sitting closer to the whiteboard need to look sideways at the whiteboard. In sum, the effect of absolute seating position on student engagement is minimal in this research.

\subsection{Relationship Between Seating Behaviours and Physiological Synchrony}
\label{sec:result_2}


As discussed in Section \ref{sec:related_sensing}, various physiological signals have been explored to infer learning engagement and affective states. From a physiological point of view, time synchronisation occurs when the physiological processes of two or more individuals are correlated with each other \cite{palumbo2017interpersonal}. Similar simultaneous changes in people's physiological signals (e.g. EDA) can provide information about cognition load or task difficulty \cite{malmberg2019we, monster2016physiological}. Recently, the physiological synchrony between individuals has been widely used in educational settings {and has become an effective way to infer group or individual engagement in an activity.}
{In particular, Gashi et al. \cite{gashi2019using} found that the engaged audiences exhibited higher levels of physiological synchrony with the presenter, which can be extended to students as the audience and the teacher as the presenter. It has also been found that the most effective teaching happens when there is a synchrony between the students and the teacher \cite{gashi2018using}.
Therefore, in this research, physiological synchrony is used to represent the similarity in learning engagement, i.e. the higher the level of physiological synchrony, the greater the similarity in student engagement between individuals. }

\subsubsection{{Data Preprocessing.}}
{The physiological synchrony was derived for each pair of students from the beginning until the end of each class. We then applied the following data prepossessing methods to the EDA signals to remove noises in the data (e.g. flat responses, movement artefacts and quantisation errors). First, we removed the incomplete data that was gathered throughout the class and discarded signals containing many movement artefacts or flat responses. Second, similar to \cite{gashi2018using,gao2020n}, a median filter with a window of five seconds was applied to the EDA signals, which reduced the artefacts but preserved typical EDA edges. Third, we decomposed the EDA signals into two parts: \textit{tonic} and \textit{phasic} \cite{boucsein2012electrodermal,cacioppo2007handbook}}. The \textit{tonic} component changes slowly and reflects the general sweat level influenced by the body or environmental temperatures. The \textit{phasic} component indicates rapid changes related to  external stimuli. The EDA signals were decomposed using \textit{cvxEDA} \cite{greco2015cvxeda} with the convex optimisation methods. Finally, we normalised the original values of the EDA signals to the range [0, 1] to make the individual signals comparable.

\begin{table}
\caption{Summary of the Pearson rank correlation results between proximity of seating and physiological synchrony across the courses. The asterisks indicate the statistically significant results:  *p < 0.05, **p < 0.01, ***p < 0.004.}
\begin{tabular}{@{}lllllll@{}}
\toprule
                     & All    & Maths  & Language & English & Politics & Science \\ \midrule
\textit{EDA\_mixed}  & {\textbf{-0.13*}} & {-0.14}  & {\textbf{-0.32***}} & {-0.02}   & {-0.11}    & {-0.01}   \\
\textit{EDA\_tonic}  & {\textbf{-0.12*}} & {\textbf{-0.24*}} & {\textbf{-0.27**}}  & {-0.05}   & {0.01}     & {0.02}    \\
\textit{EDA\_phasic} & {-0.02}  & {0.09}   & {-0.08}    & {0.01}    & {-0.21}    & {-0.09}  \\ \bottomrule
\end{tabular}
\label{tab:summaryofPS}
\end{table}

\subsubsection{{Correlation Results.}}
After the data preprocessing, the physiological synchrony was quantified using the students' normalised EDA signals. 
Based on prior research \cite{hernandez2014using, gashi2019using, malmberg2019we}, we adopted one of the most popular methods to represent physiological synchrony \textit{Pearson product-moment correlation coefficient}, which measures the linear dependence between two signals. {P-values were tested against both the $\alpha$ = 0.05 and the corrected threshold ${\alpha}_c = \frac{\alpha}{n} = 0.004$, where \textit{n} = 12. The latter is known as Bonferroni correction \cite{napierala2012bonferroni}, which is applied when \textit{n} multiple statistical tests are performed simultaneously}. We then analysed the relationship between the proximity of students and physiological synchrony using 368 pairs of instances, including the unique combination of two participants' IDs and a class ID. The number of physiological synchrony values (368 pairs) is much lower than the perceived student engagement (1,123 pairs) for the following reasons: (1) Some students reported their perceived engagement but  did not wear the E4 wristbands during the class. (2) Some students wore the E4 wristbands during the class, but the signals were not recorded for the entire class owing to the battery running flat or the wristband turning off accidentally. (3) The quality of some E4 signals was too low (e.g. too many flat responses), and these signals were removed during the preprocessing stage.

Then, we ran the correlation analysis separately on the mixed EDA signals, tonic EDA signals and phasic EDA signals. {Prior to the Bonferroni correction, there was a significant negative correlation between seating proximity and physiological synchrony (corr = -0.12 and p = 0.03). These results suggest that students sitting close together tended to experience higher physiological synchrony. However, after Bonferroni correction, this correlation was not significant. Therefore, it should not be considered as conclusive. One potential reason for this lack of significance is that we did not account for the impact of different courses during the correlation analysis. }

\subsubsection{{Impact of Different Courses.}}

Next, we explored the correlation results across the courses. We only focused on the main courses: \textit{Maths}, \textit{Language}, \textit{English}, \textit{Politics} and \textit{Science}. The other courses, such as \textit{Form}, \textit{Chapel} and \textit{Health} were not considered owing to the limited number of physiological signals. Table \ref{tab:summaryofPS} summarises the results of the \textit{Pearson} rank correlation between the seating distance and physiological synchrony. Before applying Bonferroni correction, seating proximity is only highly correlated with physiological synchrony in the \textit{Language} class (EDA\_mixed: corr = -0.32, p = 0.002, EDA\_tonic: corr = -0.27, p = 0.009) and \textit{Maths} class (EDA\_tonic: corr = -0.24, p = 0.03), and no other significant correlation was found. These results are interesting, and one possible reason for the results is that the physiological synchrony of students can be depicted more accurately in \textit{Maths} or \textit{Language} classes than in other classes, or students who sit close together  are more likely to have physiological synchrony in \textit{Maths} and \textit{Language} classes. 

In addition, we observed that statistically significant correlations were mainly reflected in the mixed EDA and the tonic EDA signals. No significant correlation was found based on the phasic EDA signals. A possible reason is that the phasic EDA signals are related to fine-grained responses to internal and external stimuli, which are usually different among students. Despite the small variations, we assume that the calculated physiological synchrony reflects general changes in student engagement during a class. Therefore, it can be seen that calculating physiological synchrony based on the tonic and mixed EDA signal works well, which may be because tonic signals indicate general arousal and learning engagement in the whole class. Since the mixed signal includes both the tonic and phasic signals, it reflects general engagement and preserves some fine-grained information from the phasic component \cite{gashi2019using}. 

\begin{table}
\caption{Description of the features computed for {electrodermal activity signals.}}
\begin{tabular}{ll}
\toprule
\textit{Feature name}   & \textit{Description of features}                                       \\ \midrule
{eda/scl/scr\_avg}   & {Average value for the EDA, SCL, SCR}                          \\
                      {eda/scl/scr\_std}   & {Standard deviation for the EDA, SCL, SCR}                     \\
                      {eda/scl/scr\_n\_}  & {Number of peaks for the EDA, SCL, SCR}                        \\
                     {eda/scl/scr\_a\_p}  & {Mean of peak amplitude for the EDA, SCL, SCR}    \\
                      {eda/scl/scr\_auc}   & {Area under the curve of the EDA, SCL, SCR}                    \\
                      {scr\_frequency} & {The frequency of phasic increases in skin conductance} \\ 
                      num\_arouse             & Number of arousing moments during the class                            \\
                      ratio\_arouse           & Ratio of arousing and unarousing moments                               \\
                      level$_k$              
                      &Ratio of the number of level$_k$ and the length of $S_k$ \\
                      eda/tonic/phasic\_pcct  & Pearson correlation coefficient with teacher                           \\
                      eda/tonic/phasic\_pccs & Pearson correlation coefficient with average value of students          \\  \bottomrule
\end{tabular}
\label{tab:feature}
\end{table}

{When Bonferroni correction was taken into account, the correlation between seating proximity and physiological synchrony was only  significant when computed using mixed EDA signals in \textit{Language} classes. The results in other courses only exhibited loose statistical significance after applying Bonferroni correction. Since we did not have a prior hypothesis for different courses, the significance before Bonferroni correction may suggest a hypothesis for future exploration.}

\subsection{Relationship Between Seating Behaviours and Physiological Arousal}
\label{sec:result_3}

As suggested in {prior research \cite{malmberg2019we,palumbo2017interpersonal,critchley2013interaction}}, physiological arousal and synchrony are both regarded as effective ways to infer people's mental activity and cognitive load. {Physiological arousal is an activity of the sympathetic nervous system, and it can be measured using EDA signals. }
Measuring physiological arousal is a useful way to understand people's emotional and cognitive processes. In general, increases in arousal are related to cognitive demand \cite{fairclough2005influence}, attention level \cite{sharot2004arousal} and learning engagement \cite{gao2020n}. In this research, we extracted some widely used features of EDA signals from previous research \cite{gashi2019using,gao2020n} to represent the general student engagement level.



\subsubsection{{Extracted Features.}}
Table \ref{tab:feature} lists {the extracted features from EDA signals based on three categories \cite{gashi2019using}, namely, general engagement, momentary engagement and synchrony:} (1) General engagement reflects the overall changes in engagement during a class. In this research, we adopted the features proposed in the literature \cite{gashi2019using, colomer2016comparison, gao2020n, babaei2021critique}, including the average, standard deviation, number of peaks, average value of peak amplitude and area under the curve, which were calculated from {EDA, SCL and SCR signals.} (2) Momentary engagement features identify evident increments in physiological arousal \cite{cain2016measuring}, {including the SCL frequency}, number of arousing moments, ratio of arousing and non-arousing moments and the ratio of the number of ${level}_k$ and $S_k$ as suggested in \cite{gashi2019using}. (3) For each student, we computed the synchrony features, such as \textit{Pearson} correlation coefficient {with the teacher and the average synchrony of students \cite{gao2020n}.}

\begin{figure}[b]
    \centering
    \includegraphics[width=0.45\textwidth]{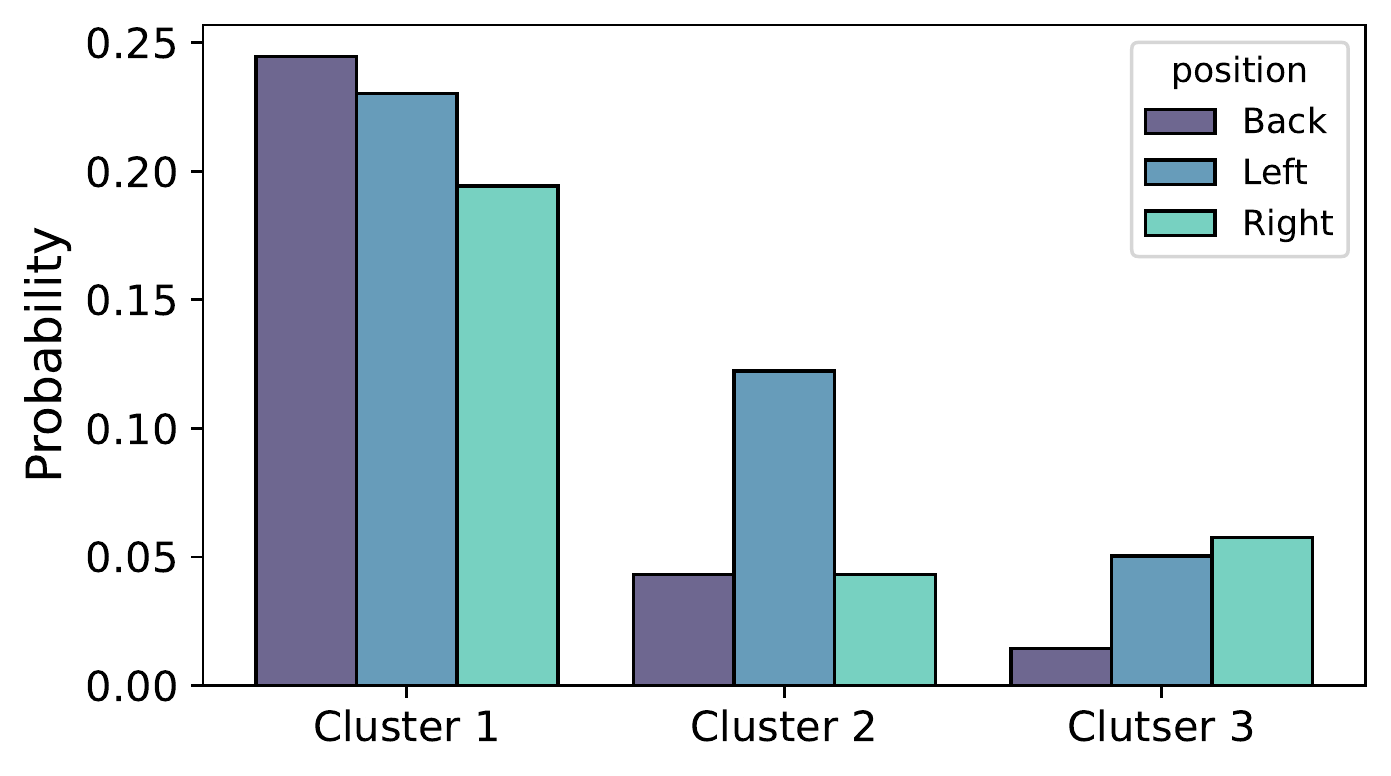}
    \caption{{The distribution of seating preferences between student groups with different physiological arousal.}}
    \label{fig:cluster_result}
\end{figure}

\subsubsection{Group-level Seating Experience.}
\label{sec:group-level}
Unlike using physiological synchrony to identify similarities in student engagement, it is difficult to intuitively compare physiological arousal {between individuals because of the numerous features that represent physiological arousal} (see Table \ref{tab:feature}). Therefore, we considered applying the clustering technique to divide the students into groups, which helped us compare patterns in physiological arousal and understand group-level student behaviours. The groups were built based on the extracted features from all EDA session logs of signal traces.

In the clustering stage, the \textit{k-means} algorithm \cite{kmeans} was adopted for clustering the features extracted from EDA signals. First, we normalised all extracted features to eliminate extreme values and ensure high-quality clusters were generated, {which was an essential step because the default \textit{Euclidean} distance metric is very sensitive to changes in the differences} \cite{virmani2015normalization}. Second, we applied the \textit{k-means} algorithm to the extracted features indicating physiological arousal and identified clusters of students. The \textit{Elbow} and \textit{Silhouette} methods \cite{lleti2004selecting} were used together to find the optimal numbers, \textit{k}, and we found that \textit{k} = 3. Next, we analysed the statistical characteristics (e.g. average value) of each clustered group and calculated students' self-reported responses in terms of seating behaviours and learning engagement.

After the clustering, we analysed the distribution of seating preferences {(i.e. back, left and right) among different clusters (see Figure \ref{fig:cluster_result})}. The clusters were divided based on the similarities in physiological arousal features, and the seating preferences were calculated based on self-report responses, as introduced in Section \ref{sec:overview}. {To determine whether the seating preferences and clusters of physiological arousal are likely to be related or not, we adopted the \textit{Pearson's chi-squared test of independence}  \cite{plackett1983karl} on above two variables, where ${\chi}^2 = 15.908$, degrees of freedom = 4, p-value = 0.003. Since the p-value was lower than 0.05, we rejected the null hypothesis and revealed that the relation between the clusters of physiological arousal and seating preferences were significant.} Figure \ref{fig:cluster_result} displayed that the students in \textit{Cluster 2} tended to sit on the left, while the students in \textit{Cluster 3} seldom sat at the back of the classroom. The students in \textit{Cluster 1} were almost equally likely to sit in any position, but they had a slightly greater tendency to sit at the back of the classroom. 



\begin{figure}
	\centering
	\subfigure[Long classes\label{fig:longclass}]{\includegraphics[width=0.45\textwidth]{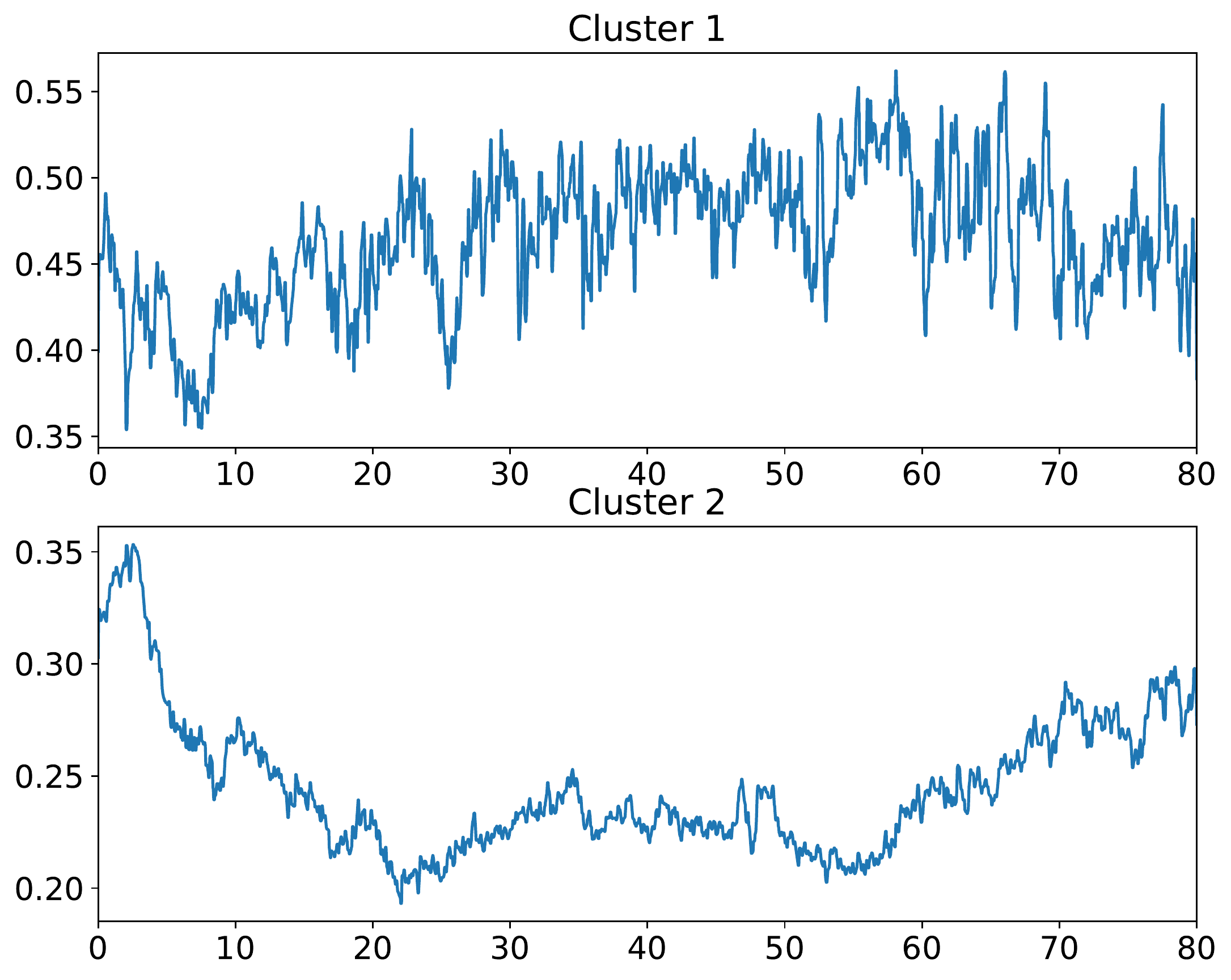}}
	\hspace{0.2cm}
	\subfigure[Short classes\label{fig:shortclass}]{\includegraphics[width=0.45\textwidth]{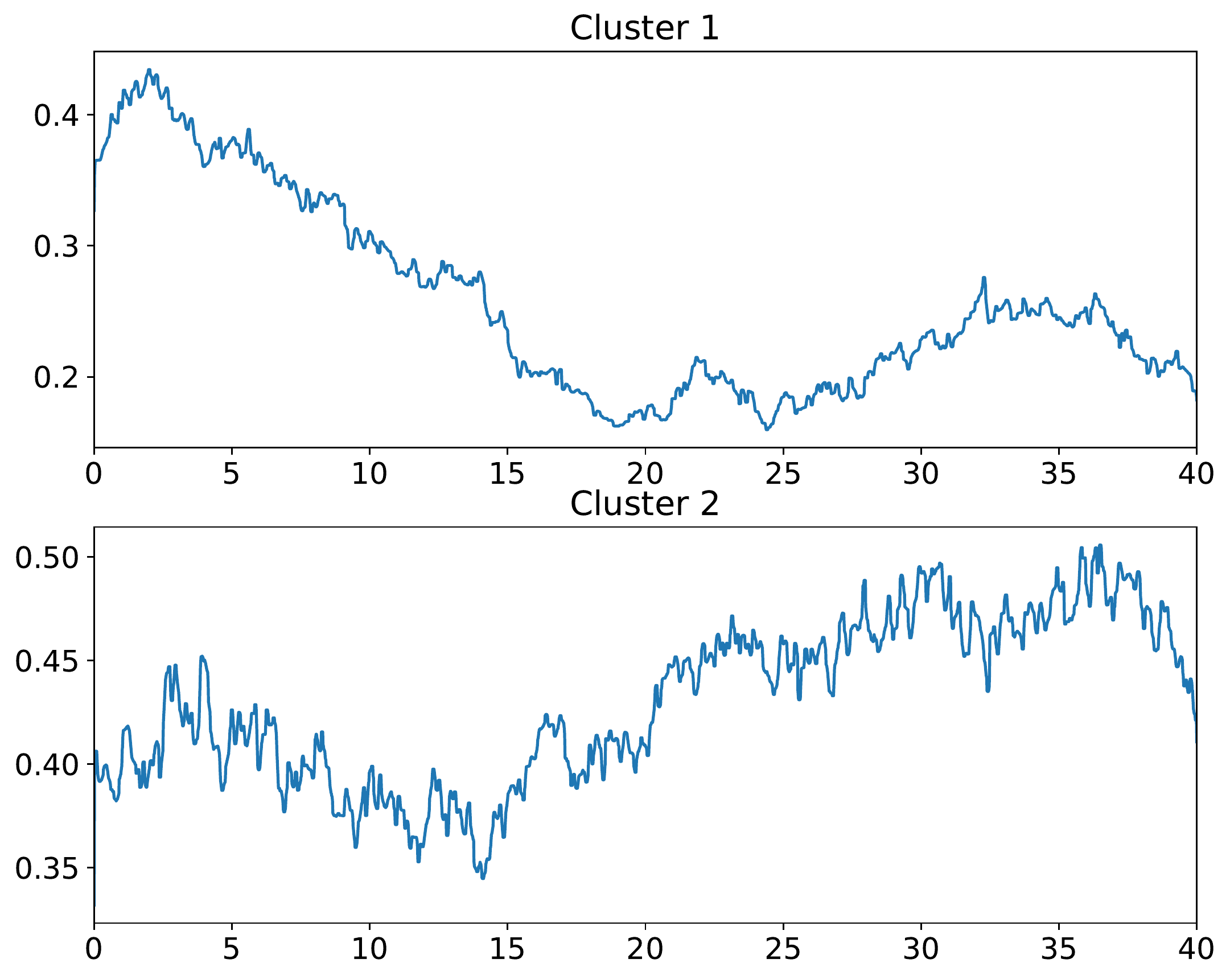}}
    \caption{{The clustering results of EDA signals for long and short classes.}}
    \label{fig:long_short_class}
\end{figure}

\begin{table}
\caption{{The learning engagement of student groups with different courses and class lengths.}}
\begin{tabular}{@{}llllll@{}}
\toprule
                  & \textit{Language} & \textit{Maths}       & \textit{English} & \textit{Politics} & \textit{Science} \\ \midrule
\textit{Cluster 1} (short) & {\textbf{3.18 (0.87)} }       & {\textbf{3.34 (0.82)}} & {3.81 (0.70)}      & {3.12 (0.75)}        & N/A      \\
\textit{Cluster 2} (short) & {\textbf{3.41 (0.90)}}        & {\textbf{3.73 (0.78)}} & {3.78 (0.70)}      & {3.32 (0.84)}       & N/A \\
\textit{Cluster 1} (long)  & {3.53 (0.81)}        & {3.46 (0.83)}          & {\textbf{3.72 (0.68)} }     & {3.53 (0.75)}       & {3.58 (0.80) }     \\
\textit{Cluster 2} (long)  & {3.70 (0.75) }       & {3.45 (0.93)  }        & {\textbf{3.93 (0.72)}}       & {3.41 (0.73) }      & {3.61 (0.68) }     \\ \bottomrule
\end{tabular}
\label{tab:subject}
\end{table}

\subsubsection{{Impact of Course and Class Length}}
Next, we explored how physiological arousal is related to perceived student engagement. Two factors may affect the clustering results: the length of the class and the course itself. Figure \ref{fig:longclass} and Figure \ref{fig:shortclass} show the average value of the EDA signals in different clusters for long and short classes. First, we divided the EDA signals based on class length: long classes (80 min) and short classes (40 min). Then we identified that the optimal number of clustering was 2. {For short classes, the statistical characteristics of self-reported overall engagement  in \textit{Cluster 1} were mean = 3.64, std = 0.82, and in \textit{Cluster 2} were mean = 3.41, std = 0.73. Results from t-test \cite{kim2015t} indicated that two clusters were significantly different in perceived emotional engagement, where t-statistic = -2.01, p-value = 0.04. From Figure \ref{fig:shortclass}, we found that there were several peaks in the EDA signals in \textit{Cluster 1}, indicating higher emotion arousal, as discussed in the literature \cite{boucsein2012electrodermal}. The perceived engagement scores also supported the clustering results of the EDA signals. However, for long classes, we could not find significant differences in any dimension of engagement in the two clusters (p-value > 0.05). }

{We then calculated the learning engagement of student groups in different courses. We did not consider courses with fewer than 100 signal traces and only focused on the courses with sufficient data (see Table \ref{tab: wearable signals}), i.e. \textit{Language}, \textit{Maths}, \textit{English}, \textit{Politics} and \textit{Science}. The student groups were then generated based on the clustering results of the physiological arousal features, and the optimal number of clusters was 2 for the above courses. Table ~\ref{tab:subject} shows the learning engagement of the student groups in different courses and the length of classes. The learning engagement in short classes for \textit{Science} is \textit{N/A} because all \textit{Science} classes were the same length (i.e. long). Table ~\ref{tab:subject} shows a significant difference in learning engagement in the two student groups for short classes in the \textit{Maths} courses (average learning engagement of \textit{Cluster 1} = 3.34 and \textit{Cluster} 2 = 3.73. Similar phenomena occurred in the short classes for \textit{Language} courses and long classes for \textit{English} courses. The potential reasons for not finding an obvious difference in the learning engagement in \textit{Politics} and \textit{Science} courses are as follows: (1) Students tended to have similar learning engagement in the above two courses than the other courses, making it difficult to determine to impact of the courses on the student groups. (2) There was a lack of physiological signals or low-quality physiological signals in these classes. Students need to do experiments in \textit{Science} classes, resulting in a high chance of signal noise, such as poor contact between the sensors and the skin.}

\section{Implications and Limitations}
\label{sec:discussion}

\textit{Implications}.
Investigating how student seating experience relate to learning engagement can help educators and policymakers improve student engagement in a variety of ways, such as providing better seating arrangements, organising more effective study groups and addressing problems such as poor academic achievement, student boredom and disaffection. Our study adds quantifiable evidence on the relationship between seating location and student engagement. The use of physiological arousal and physiological synchrony provides an effective method of understanding student engagement in real-time, with less burden on students than traditional questionnaires. 

This research shows the possibilities of developing an intelligent seating recommendation system integrated with wearable devices to optimise student engagement in the classroom. The system could analyse student engagement in real-time and provide recommendations on the best or alternative seating positions in the classroom. By understanding student seating preferences, the system may provide learning tips for students, e.g. if a student's engagement decreases as a result of their seating choices, the system could suggest changing to another seat or increasing participation in class activities. The system will also benefit teachers by enabling them to manage students more effectively, e.g. facilitating student learning by monitoring students' behavioral patterns in the classroom. Furthermore, capturing student engagement and seating experience may aid in curriculum design, thereby increasing teaching efficiency and improving teaching style. 


\textit{Limitations}.
This study has some limitations that need to be overcome in future studies. First, although efforts were made to ensure that the modified five-item questionnaire captured the same constructs as the original questionnaire, validating the modified version was challenging due to the limited number of participants and budgetary constraints. It is also not feasible to adopt the traditional questionnaires (e.g. EvsD \cite{skinner2008engagement}) in education, as they often contain dozens of questions which would create a burden on students if they needed to be filled out after each class. However, the risk of a lack of validity of the questionnaire is minimal because we only slightly adapted it to suit the high school context, and the adapted questions were almost identical to those used in widely accepted questionnaires \cite{skinner2008engagement} in the educational context. Second, the study into group-wise experiences was limited to studying student peers and groups based on clustering. Future research could investigate different psychological patterns within and between student groups. 

Third, the accuracy of the EDA signals measured by the E4 wristbands is limited. Menghini et al. \cite{menghini2019stressing} evaluated the accuracy of the E4 wristbands compared with the gold-standard sensors and found that similar accuracy could not be achieved using EDA signals from the wrists and fingers. A promising solution to improve the accuracy of the E4 may be lead wire extension, which would allow EDA recordings to be taken from the finger or palm rather than the wrist, thereby eliminating any potential site difference. However, in this study, the E4 wristbands were the best choice to obtain data without disturbing students in the classroom.

Furthermore, since participants were required to wear wristbands during school, sweat accumulation may have affected our results, especially given the hours of recording and the use of dry electrodes. In our data collection, it’s challenging to control the factors related to the learning environment such as the humidity and temperature. However, the students' classrooms and teaching buildings were equipped with central air conditioners, so that the indoor temperature was not too high/too low, and the students did not sweat too much. After each class (40 minutes or 80 minutes), students would take breaks and walk, which is conducive to the evaporation of sweat from their body surface. When analyzing EDA signals, the tonic signal changes slowly and reflects the general sweat level influenced by the body or environment temperatures, while the phasic component indicates the rapid changes related to the external stimuli. Despite the influence of sweat accumulation, the phasic EDA may still indicate student engagement.

Last but not least, in our research, some procedures for collecting and processing EDA data may not have strictly followed the best standards of EDA practice (e.g. caffeine control, medication control, counter-balancing of external factors) as suggested by Babaei et al. \cite{babaei2021critique}, thus threatening the validity of our results to some extent. However, with our data collection spanning four weeks, it was not feasible to control for various external factors without burdening students and affecting their daily learning. In addition, due to the relatively low quality of the collected EDA signals compared to laboratory studies, it was challenging for us to apply standard settings to signal processing in the recommended way \cite{babaei2021critique}. Therefore, we have followed some EDA signal processing methods (e.g. applying a median filter with a window of five seconds) that have proven effective in similar data collection environments \cite{di2018engagement,gao2020n}. Additionally, to compute features from EDA signals, more useful features representing student engagement could be explored, and some features that may be less relevant to arousal could be reduced to improve the performance of clustering in Section \ref{sec:group-level}. In the future, more community standards of EDA practices could be explored for field studies to improve the validity of the research.

\section{Conclusion}
\label{sec:conclusion}
In this research, we explored how student seating experiences are related to their learning engagement by understanding their physiological and behavioural patterns during a four-week data collection. The results showed that the individual and group-wise classroom seating experience is statistically correlated with {both perceived engagement and physiologically measured student engagement (physiological synchrony and physiological arousal).} We found that students who sat close together were more likely to have similar learning engagement and higher physiological synchrony than students who sat far apart. These findings open up opportunities on exploring the implications of seating arrangements and recommendations to improve student engagement in classrooms.  

\section*{Acknowledgments}

This research is supported by the Australian Government through the Australian Research Council's Linkage Projects funding scheme (project LP150100246). This paper is also a contribution to the IEA EBC Annex 79.

\bibliographystyle{ACM-Reference-Format}
\bibliography{nanref}

\end{document}